\documentstyle[preprint,aps,prl]{revtex}

\newcommand{\be}{\begin{equation}}
\newcommand{\ee}{\end{equation}}
\newcommand{\ba}{\begin{eqnarray}}
\newcommand{\ea}{\end{eqnarray}}
\newcommand{\ket}[1]{| {#1} \rangle}
\newcommand{\bra}[1]{\langle {#1} |}

\newcommand{\mpi}{m_\pi}

\def\lsim{\mathrel{\rlap{\lower4pt\hbox{\hskip1pt$\sim$}}
    \raise1pt\hbox{$<$}}}     %less than or approx. symbol
\def\gsim{\mathrel{\rlap{\lower4pt\hbox{\hskip1pt$\sim$}}
    \raise1pt\hbox{$>$}}}     %greater than or approx. symbol

\begin{document}

\draft

\title{Medium Modifications of the Rho Meson at CERN/SPS Energies}

\author{R. Rapp$^{1}$,  G. Chanfray and J. Wambach$^{1,2}$}

\address{
Institut de Physique Nucl\'eaires,
43 Avenue de 11 Novembre 1918,
69622 Villeurbanne Cedex, France \\
1) also at: Institut f\"ur Kernphysik, Forschungszentrum
J\"ulich GmbH, D-52425 J\"ulich, Germany \\
2) Permanent address: Department of Physics, University
of Illinois at Urbana-Champaign, 1110 W. Green St.,
Urbana, IL 61801, USA}

%\date{23.08.95}

\maketitle

\begin{abstract}
Rho meson propagation in hot hadronic matter is studied in
a model with coupling to $\pi\pi$ states. Medium modifications
are induced by a change
of the pion dispersion relation through collisions with nucleons and
$\Delta's$ in the fireball. Maintaining gauge invariance dilepton
production is calculated and compared to the recent data of the CERES
collaboration in central S+Au collisions at 200 GeV/u. The observed
enhancement of the rate below the rho meson mass can be
largely accounted for.
\end{abstract}


% body of paper here
In central heavy-ion reactions at ultrarelativistic energies hot and
dense hadronic matter is formed at the early stages of the collision.
 From such experiments one hopes to infer properties of mesons and
baryons in the vicinity of the chiral phase transition. Because of its
electromagnetic origin, lepton pair emission is an ideal probe. Once
produced, the dilepton will decouple from the strongly interacting
particles and thus carry undistorted
information of the dynamical properties of the
hot nuclear system. For this reason the recent low-mass
dilepton spectra
measured by the CERES collaboration in p+Be,
p+Au and S+Au~\cite{CERES} collisions at
450 GeV/u and 200 GeV/u, respectively, have produced considerable
interest.
While the proton data are well
described by known decays from a hadronic fireball an excess of
dielectrons with invariant masses of 200 MeV to 1 GeV
has been observed in the S+Au case
and has been attributed to $\pi^+\pi^-$ annihilation from the pionic
component of the hadron gas~\cite{CERES}. When this process is included
in the
description of the data the excess of strength between the
$2 \mpi$-vacuum threshold and the rho peak is still unexplained. This
led the authors of ref.~\cite{LKB} to the conclusion that one is seeing
a
signal of partial restoration of chiral symmetry through a decrease of
vector meson masses (here especially the rho meson) near the phase
boundary. Since the rho meson strongly
couples to the $\pi^+\pi^-$ channel
this effect may also be explained more conventionally, however, namely
by modifications of the pion propagation in the presence of nucleons and
isobars~\cite{KoPr,AsKo,ChSc,HeFN,Henn,RaWa2,ARCSW}.

To study the dilepton emission from $\pi^+\pi^-$ annihilation we
invoke the vector-dominance model (VDM) in which the electromagnetic
current is related to the third
isospin component of the rho meson field by the
field current identity $J^\mu=(m_\rho^2/g)\rho^\mu_3$ where $m_\rho$ and
$g$ denote the rho meson mass and the universal VDM coupling constant,
respectively. In terms of the retarded (thermal) rho-meson propagator
\be
D^{\mu\nu}_\rho(q;T)=-i\int d^4x e^{-iqx}\sum_i{e^{-\beta E_i}\over Z}
\bra{i}[\rho^\mu_3(x),\rho^\nu_3(0)]\ket{i}\Theta(x_0)
\ee
($\beta=1/T$) the dilepton emission rate per unit volume,
$R$=$dN_{l^+l^-}/d^4x$, is given by
\be
{dN_{l^+l^-}\over d^4xd^4q}={\alpha^2\over 3\pi^2 M^2}H(q;T)
\ee
where $q=p_++p_-=(q_0,\vec q)$ denotes the total momentum for (massless)
dileptons, $M^2\equiv q^2_\mu > 0$ is the invariant mass and the
hadronic piece $H(q;T)$ is given by
\be
H(q;T)=f^\rho(q_0,T)((m_\rho^{(0)})^4/\pi g^2)g_{\mu\nu}{\rm Im}
D^{\mu\nu}_\rho(q;T)\
\ee
($f^\rho(q_0,T)=(e^{q_0/T}-1)^{-1}$).
To reduce the numerical complexity in evaluating $D^{\mu\nu}_\rho$
we shall restrict ourselves to
back-to-back kinematics, {\it i.e.} $\vec q=0$. In that case, from gauge
invariance, only the space components contribute. Denoting
$D^{ij}_\rho(q_0,\vec q=0;T)=\delta^{ij}D_\rho(q_0;T)$ eq.~(3) then
reduces to
\be
H(q_0,\vec q=0;T)=-f^\rho(q_0,T){3(m_\rho^{(0)})^4\over\pi g^2}
{\rm Im}D_\rho(q_0;T)
\ee
where
\be
D_\rho(q_0;T)=[q_0^2-(m_\rho^{(0)})^2-\Sigma_\rho(q_0;T)]^{-1}
\ee
involving the scalar part of the rho self energy $\Sigma_\rho$.

To study medium modifications of the $\rho$ propagator we first need a
realistic model for the $\rho$ meson in free space. As is well
known~\cite{JoLe} an appropriate description is provided
by a bare pole graph renormalized by $\pi\pi$ rescattering.
This amounts to solving a Lippmann--Schwinger type equation
\be
M_{\pi\pi}(E,q_1,q_2)  =  V_{\pi\pi}(E,q_1,q_2)
+ \int\limits_{0}^{\infty}dk \ k^2 \  V_{\pi\pi}(E,q_1,k)
 \   G_{\pi\pi}^0(E,k) \ M_{\pi\pi}(E,k,q_2)  \ ,
\label{eq:Mamp}
\ee
where $E$ is the starting energy of the pion pair, $q_1$ and $q_2$
denote the moduli of the pion three-momenta in the cms frame and
$G^0_{\pi\pi}(E,k)$ is the free two-pion propagator:
\be
G_{\pi\pi}^0(E,k)=\frac{1}{\omega_k} \  \frac{1}{E^2-4\omega_k^2+i\eta}
 \ ; \quad \omega_k^2=m_\pi^2+k^2
 \ .
\ee
The pseudopotential generated by the  s--channel $\rho$ pole graph is
given by
\be
V_{\pi\pi}(E,q_1,q_2)=v(q_1)D_{\rho}^0(E) v(q_2)  \  ,
\ee
where
\ba
D_{\rho}^0(E) & = & [E^2-(m_{\rho}^{(0)})^2]^{-1}
\nonumber \\
v(q) & = & \sqrt{\frac{2}{3}}
\frac{g}{2\pi} \ F(q)
\ea
are the bare $\rho$ propagator and the vertex functions, respectively.
The hadronic form factor is chosen to be of dipole form normalized
to one at the physical resonance energy $m_\rho$=770~MeV:
\be
F(q)=\biggl (\frac{2\Lambda_\rho^2+m_\rho^2}{2\Lambda_\rho^2
+4\omega_q^2} \biggr)^2
 \ .
\ee
The parameters $g$, $\Lambda_\rho$ and $m_\rho^{(0)}$
are fitted to the experimental p-wave $\pi\pi$ phase shift
and the pion electromagnetic form factor
\ba
|F_\pi(E)|^2 & = & \frac{(m_\rho^{(0)})^4}{(E^2-(m_\rho^{(0)})^2-
Re\Sigma_\rho^0(E))^2+Im\Sigma_\rho^0(E)^2} \nonumber\\
& \equiv & (m_\rho^{(0)})^4 \ |D_\rho^0(E)|^2
\ea
with
\ba
\Sigma_\rho^0(E) & = & \bar{\Sigma}_\rho^0(E)-\bar{\Sigma}_\rho^0(0)
 \ , \nonumber\\
\bar{\Sigma}_\rho^0(E) & = & \int  \ k^2 \ dk \ v(k)^2 \
G_{\pi\pi}^0(E,k)  \  .
\ea
The subtraction for $\Sigma_\rho^0$ at E=0 is necessary to ensure the
normalization $F_\pi(0)=1$.  We obtain a satisfactory fit with
$g^2/4\pi$=2.7, $\Lambda_\rho$=3.1~GeV and $m_\rho^{(0)}$=829~MeV
as indicated in fig.~1.

A proper calculation of the in-medium $\rho$ propagator
\be
D_\rho(E)=[E^2-(m_\rho^{(0)})^2-\Sigma_\rho(E)]^{-1} \ ,
\ee
must include constraints from gauge invariance. A detailed derivation
for the self energy in homogeneous matter is
given in refs.~\cite{ChSc,HeFN},
and we do not repeat it here. The result can be expressed
in terms of the spin--isospin longitudinal and
transverse response functions at a given density and temperature
\ba
\Pi_L(k_0,k) & = & (k_0^2-k^2-m_\pi^2) \ \tilde\Pi^0(k_0,k)
\ D_L(k_0,k) \ ,
\nonumber\\
\Pi_T(k_0,k) & = & (k_0^2-k^2-m_\rho^2) \ \tilde\Pi^0(k_0,k)
\ D_T(k_0,k) \
\ea
where $D_L$ and $D_T$ are the longitudinal and transverse propagators
which are expressed in terms of the pion self energy as
\ba
D_L(k_0,k) & = & \lbrack k_0^2-k^2-m_\pi^2-
\Sigma_\pi(k_0,k) \rbrack ^{-1} \ ,  \nonumber\\
D_T(k_0,k) & = & \lbrack k_0^2-k^2-m_\rho^2-
C_\rho\Sigma_\pi(k_0,k) \rbrack ^{-1} \ , \nonumber\\
\tilde\Pi^0(k_0,k) & = & \frac{1}{k^2} \Sigma_\pi(k_0,k) \ .
\ea
Obviously $D_L$ is equal to single pion propagator $D_\pi$.
Making use of derivations in ref.~\cite{ChSc} the imaginary part of the
self energy for a $\rho$ meson at rest then takes the form
\ba
{\rm Im} \Sigma_\rho(q_0,\vec 0) & = & \int\limits_{0}^{\infty}
k^2 \ dk \ v(k)^2 \int\limits_0^{q_0} \frac{dk_0}{\pi}
{\rm Im} D_\pi(q_0-k_0,k)
\nonumber\\
 & &   \times
{\rm Im}  \lbrace
\alpha(q_0,k_0,k) D_\pi(k_0,k)+\frac{1}{2}  \Pi_L(k_0,k)+
\Pi_T(k_0,k) \rbrace ,
\ea
where the function $\alpha$ is given by
\ba
\alpha(q_0,k_0,k) &  =  &
   [1+\tilde\Pi^0_R(k_0,k)+\tilde\Pi^0_R(q_0-k_0,k)
 + \frac{1}{2}\tilde\Pi^0_R(k_0,k) \tilde\Pi^0_R(q_0-k_0,k)]
\nonumber\\
\tilde\Pi^0_R(k_0,k) &  = & {\rm Re} \tilde\Pi^0(k_0,k) \ .
\ea
The real part of $\Sigma_\rho$ is obtained via a dispersion integral:
\be
Re \Sigma_\rho(q_0)=-{\cal P} \int\limits_0^\infty
\frac{dE'^2}{\pi} \frac{{\rm Im} \Sigma_\rho(E')}{q_0^2-E'^2}
\frac{q_0^2}{E'^2} \ .
\ee
The factor $q_0^2/E'^2$ results from a subtraction at $q_0$=0,
which is necessary to ensure gauge invariance~\cite{ChSc}; the
imaginary part is unaffected by this subtraction since
$Im \Sigma_\rho(q_0=0)=0$.

It remains to specify the model for the single--pion
self energy $\Sigma_\pi$. We employ the standard model of p--wave
particle--hole excitations extended to finite temperature~\cite{RaWa2}.
Here the pions are dressed by $NN^{-1}$, $\Delta N^{-1}$ as well as
$N\Delta^{-1}$ and $\Delta\Delta^{-1}$ excitations, the latter appearing
as a consequence of a thermally excited $\Delta$ abundance in the gas.
One has
\be
\Sigma_{\pi}(\omega,k)= -k^2 \
\sum_\alpha \chi_{\alpha}(\omega,k)  \ ,
\ee
where the summation is performed over all excitation channels
$\alpha$=$ab^{-1}$, $a$,$b$=$N$,$\Delta$. The susceptibilities
$\chi_\alpha$ contain short-range correlations taken into account by
Landau-Migdal parameters  ${g'}_{\alpha\beta}$ such that
\ba
\chi_\alpha & = & \chi_\alpha^{(0)}-\sum_{\beta} \
\chi_\alpha^{(0)} \ {g'}_{\alpha\beta} \ \chi_\beta
 \ \nonumber\\
\chi_\alpha^{(0)}(\omega,k) & = & {\left
( \frac{f_{\pi\alpha} \  \Gamma_\pi(k)} {m_\pi}
\right )}^2  \ SI(\alpha) \ \phi_\alpha(\omega,k) \ ,
\ea
where $f_{\pi\alpha}$ denote the $\pi NN$ {\it etc}
coupling constants and
$SI(\alpha)$ is a spin--isospin factor. The form factor $\Gamma_\pi(k)=
(\Lambda_\pi^2-m_\pi^2)/(\Lambda_\pi^2+k^2)$ accounts
for the hadronic size
of the pion-baryon vertex ($\Lambda_\pi=1200$ MeV). The explicit form of
the thermal Lindhard functions, $\phi_\alpha$, including the $\Delta$
width has been given in ref.~\cite{RaWa2}. For the Migdal parameters we
take ${g'}_{\alpha\beta}=0.8$ for $\alpha\beta$=$aa^{-1}bb^{-1}$
and ${g'}_{\alpha\beta}=0.5$ for all others.
Finally, to account for a finite pion density,
we supplement eq.~(16) with a 2-pion Bose factor,
$[1+f^\pi(k_0)+f^\pi(q_0-k_0)]$, which can be derived
rigorously within a Matsubara formalism~\cite{RaWa3}.
The contribution of the gas pions to the pion self energy
$\Sigma_\pi(k_0,k)$ has been shown to be small~\cite{RaWa3} and we can
safely neglect it here.

Expressions (2)-(20) specify our model for the dilepton rate
in back-to back kinematics. Since we evaluate the $\rho$ propagator for
a homogeneous gas of nucleons and $\Delta$'s in thermal equilibrium,
the rate at a given temperature $T$ of the fireball is determined
by the nucleon and $\Delta$ abundances at that $T$. To evaluate
these abundances
we make use of the transport results of Li {\it et al.}~\cite{LKB}. It
is found that at the initial stage the total baryon density $\rho_b$ is
$3.5\rho_0$ ($\rho_0=0.16$~fm$^{-3}$) at
a temperature of 170 MeV. Including besides the nucleon and $\Delta$ all
baryon resonances with masses below 1.7 GeV as well as the lowest-lying
hyperons~\cite{LKB} and assuming chemical
equilibrium,  the chemical potential
$\mu_N=\mu_\Delta=\dots=0.448$~GeV is determined by the initial baryon
density. The corresponding nucleon and $\Delta$ densities are
1.0 $\rho_0$ each, the sum being in good agreement with the
results quoted in ref.~\cite{LKB}. The time evolution of the abundances
can be obtained from the time dependence of the temperature
given in ref.~\cite{LKB}. A reasonable fit of the transport
results is obtained with
\be
T(t)=(T^i-T^\infty) e^{-t/\tau} + T^\infty
\ee
with an initial temperature $T^i$=170~MeV, $T^\infty$=115~MeV
and $\tau$=10~fm/c. For chemical equilibrium the
time evolution of $\rho_b$
can then be calculated and the result is again in agreement with
the transport calculations~\cite{LKB}. Thus we can extract $\rho_N$ and
$\rho_\Delta$ at each time for the integration of the time history.

For a direct comparison with experiment two further points have to be
considered. The first is related to the fact that the rate is only
evaluated in back-to-back kinematics ($\vec q$=0).  Minimally this can
be corrected for by using~\cite{SKLK}
\ba
\frac{dN_{l^+l^-}}{d^4xd^4q}(q_0,\vec q) & = &\biggl
(\frac{dN_{l^+l^-}}{d^4xd^4q}
\biggr )_{\vec q=0}F(M,q,T)\nonumber\\
F(M,q,T) & = & \frac{\exp [-(\sqrt{{\vec q}^2 +M^2}-M)/T]}
{\sqrt{1+{\vec q}^2/M^2}}
\ea
which is exact for free pions. The second, more severe, point is the
finite momentum acceptance of the detector. In the CERES S+Au experiment
only dielectrons with opening angles
$\Theta_{ee}>35$~mrad and transverse
momenta $p_t>0.2$GeV are detected. While the opening angle restriction
is not severe the $p_t$-cuts have to be included properly. To do so
we statistically model $e^+e^-$ $\rho$ decays at given
temperature and invariant mass and apply the
acceptance cuts to each of the two lepton tracks~\cite{VKpr} resulting
in an acceptance function $A(M,T)$. The final dilepton yield can now
be calculated by taking into account the above modifications
and integrating over the time evolution of the fireball up to
the freeze-out time $t_f$ to obtain
\be
\frac{dN_{e^+e^-}}{d^3xdM} =
\int\limits_{0}^{t_{f}} dt \
C(M,T(t)) \ \biggl (\frac{dN_{e^+e^-}(t)}{d^4xd^4q}
\biggr )_{\vec q=0}
\ee
where
\be
C(M,T)=A(M,T)\int \frac{d^3q}{(2\pi)^3}F(M,q,T)
\ee
contains the acceptance function as well as the finite $q$
correction to the back-to-back rate (eq.~(2)). We take $t_f$=10~fm/c,
but our results are affected less than 10$\%$ when choosing
$t_f$=15fm/c. Finally we fix the
overall normalization in the measured rapidity window to the total yield
from vacuum $\rho$ decay obtained by Li {\it et al.}~\cite{LKB}.
Fig.~2 shows our final results supplemented with contributions from
free Dalitz decay of $\pi^0$'s, $\eta$'s and $\omega$'s (dashed-dotted
line), which we extracted from ref.~\cite{CERES}. We also implemented
the contributions from free $\omega$ decay ($\omega\rightarrow e^+e^-$)
according to ref.~\cite{LKB}, which,
due to the small $\omega$ width, are expected to undergo
only minor medium modifications~\cite{LKB}.
The inclusion of the medium modifications of
the $\rho$ meson leads to an enhancement of the $e^+e^-$ yield
of about a factor of 3 for invariant masses around
M$\approx$0.5~GeV/$c^2$
(full line compared to the dotted line in fig.~2).

In summary, we have presented a gauge invariant calculation
of the $\rho$ propagator at rest in a hot $\pi$N$\Delta$ gas
taking into account the full off-shell dynamics of the intermediate
pions when coupled to nucleons and $\Delta$'s. After correcting for
a finite momentum of the dilepton pair
as well as for the experimental acceptance of the CERES detector,
our calculated yield can, to a large extent, account for the
enhancement in the $e^+e^-$  spectra observed in 200~GeV/u
S+Au collisions in the CERES experiment.

\noindent

\bigskip

\noindent
{\bf Acknowledgement}:
We thank W. Cassing, M. Ericson, C.M. Ko and V. Koch for fruitful
discussions. We are especially indebted to V. Koch for providing us with
a Monte Carlo program to determine the detector acceptance in the CERES
experiment.
This work is supported in part by the National
Science Foundation under Grant No. NSF PHY94-21309.

% now the references. delete or change fake bibitem. delete next three
%   lines and directly read in your .bbl file if you use bibtex.

\newpage
%%%%%%%%%%%%%%%%%%%%%%%%%%%%%%%%%%%%%%%%%%%%%%%%%%%%%%%%%%%%%%%%%%%%%
% Figure Captions
%%%%%%%%%%%%%%%%%%%%%%%%%%%%%%%%%%%%%%%%%%%%%%%%%%%%%%%%%%%%%%%%%%%%%
\begin{center}
{\large \sl \bf Figure Captions}
\end{center}
\vspace{0.5cm}

\begin{itemize}
\item[{\bf Figure 1}:] Our fit to the p-wave $\pi\pi$ phase shifts
(upper panel) and the pion electromagnetic form factor (lower panel);
the squares in the lower panel are the values from the
Gounaris-Sakurai formula~\cite{GoSa}, which itself gives an
accurate description of the data.

\item[{\bf Figure 2:}]
Dielectron yield from free Dalitz decay (dashed-dotted line),
free Dalitz + free $\omega$ + free $\rho$ (dotted line) and
free Dalitz + free $\omega$ + in-medium $\rho$ decay (full line);
the dots are the CERES data.

\end{itemize}

% Put the label that you will use with \ref{} command in
% the braces of the \label{} command.
% Insert the column specifiers (l, r, c, d, etc.)
% in the empty braces of the
% \begin{table}
% \caption{}
% \label{}
% \end{tabular}
% \end{table}

%\begin{table}
%\squeezetable
%\caption{Kinematics for the ${}^{12}$C(e,e$^\prime$p) measurements.}
%\label{tab1}
%\begin{tabular}{llllll}
%$Q^2$ & $E$ & $E^\prime$ & $P$ & $\theta_e$ & $\theta_p$\cr
%$\hbox{(GeV/c)}^2$&GeV&GeV&GeV/c&  & \cr
%\tableline
%1.04&2.02&1.39&1.20&35.5$^\circ$&43.4$^\circ$,46.2$^\circ$,49.0
%$^\circ$,
%51.8$^\circ$,54.6$^\circ$\cr
%3.06&3.19&1.47&2.45&47.7$^\circ$&27.7$^\circ$, 30.5$^\circ$, 33.3
%$^\circ$\cr
%5.00&4.21&1.47&3.54&53.4$^\circ$&20.9$^\circ$, 22.6 $^\circ$\cr
%6.77&5.12&1.47&4.49&56.6$^\circ$&15.9$^\circ$, 17.3$^\circ$\cr
%\end{tabular}
%\end{table}

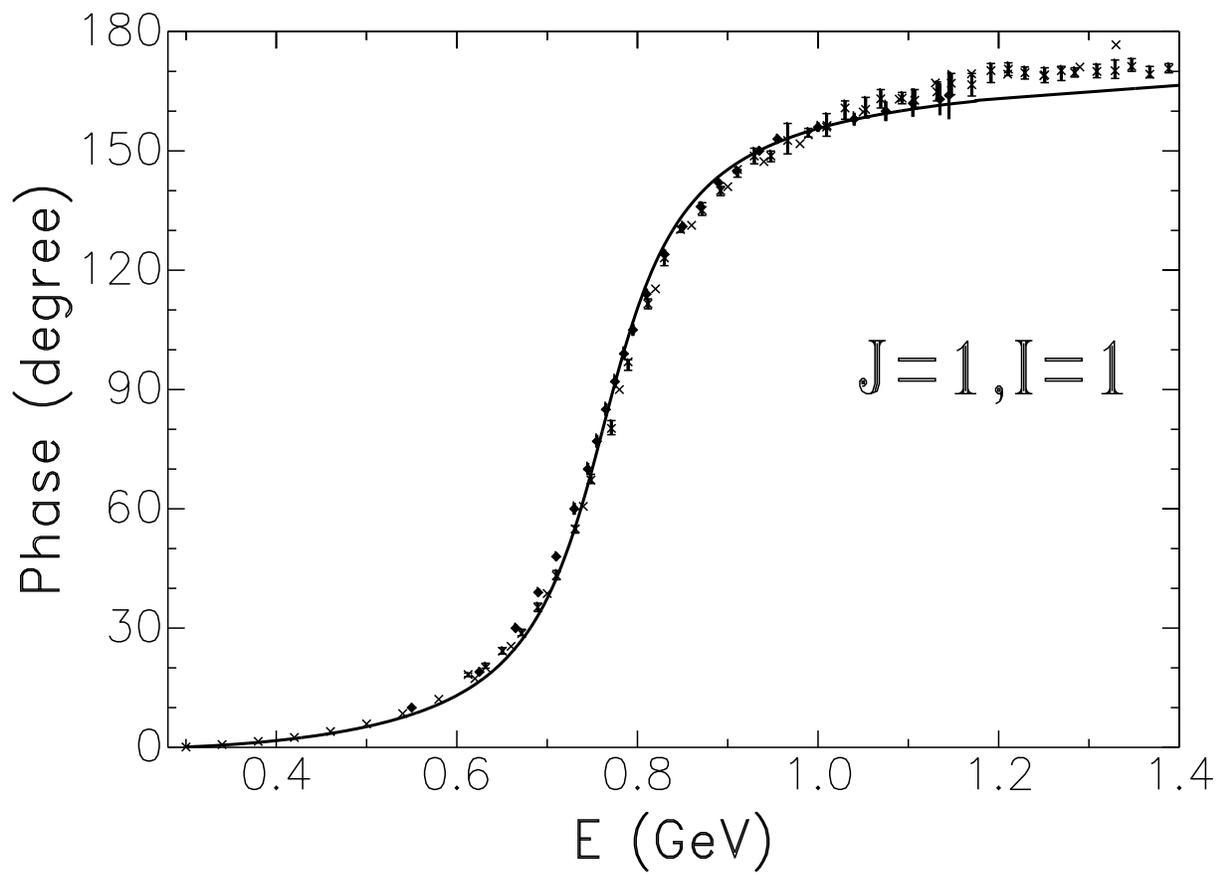
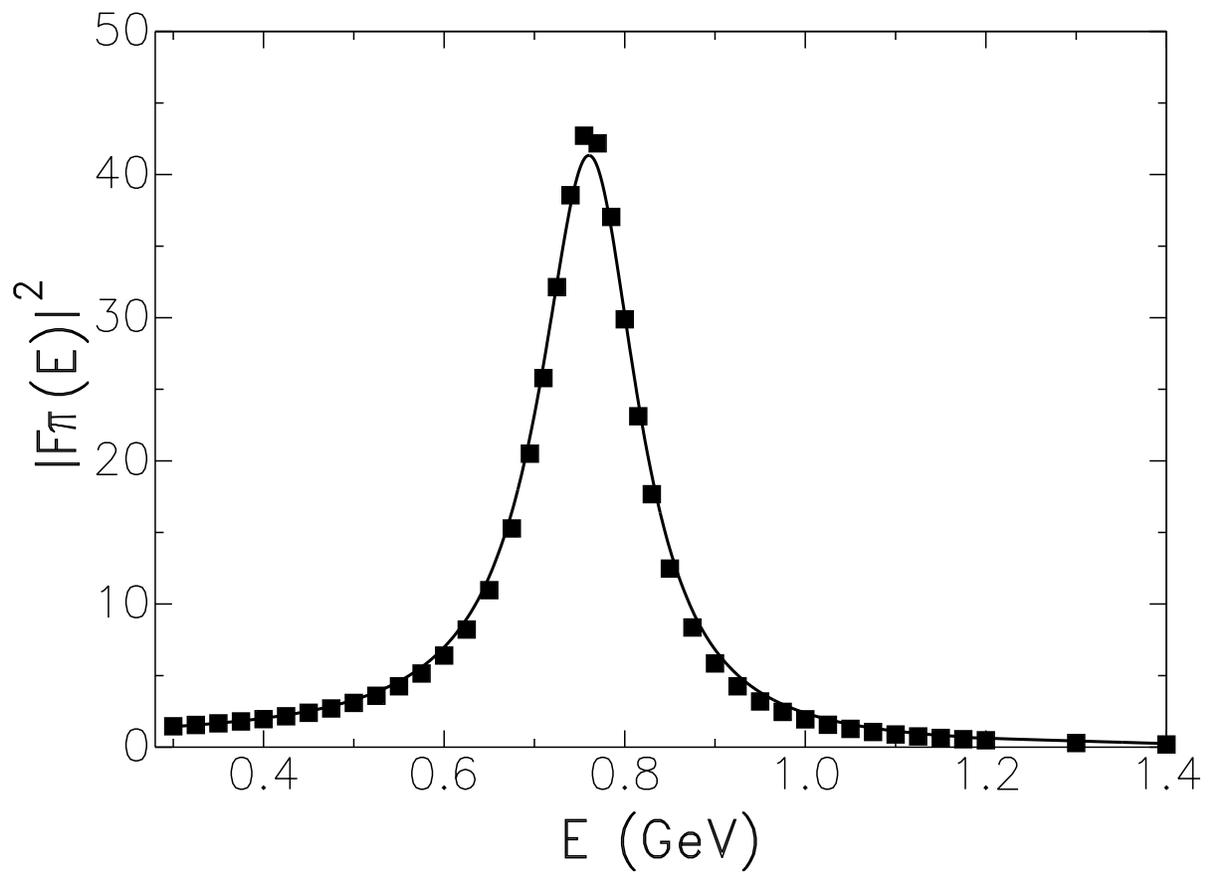
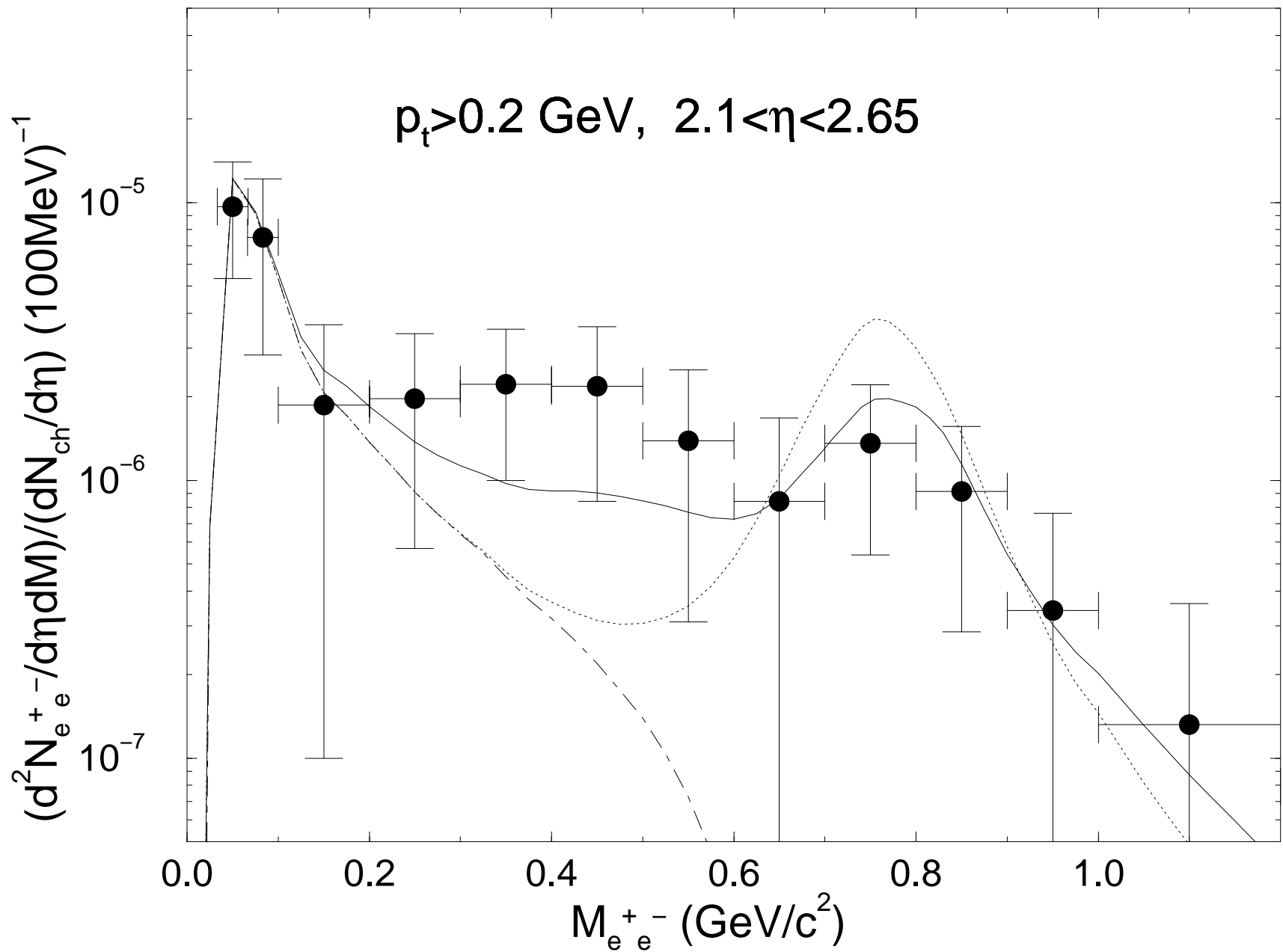
\end{document}